\documentclass{article}
\usepackage{spconf,amsmath,graphicx}
\usepackage{amssymb}
\usepackage{amsfonts}
\usepackage{graphicx}
\usepackage{pdfpages}
\usepackage{setspace}
\usepackage{multirow}
\usepackage{amsmath}
\usepackage{empheq}
\usepackage{epsfig}
\usepackage{array}
\usepackage{color}
\usepackage{subfigure}
\usepackage{xcolor}
\usepackage{amssymb}
\usepackage{url}
\usepackage{mathrsfs}
\usepackage{booktabs}
\usepackage{cite}
\usepackage{bbding}
\usepackage{upgreek}

\title{A NEURAL BEAM FILTER FOR REAL-TIME MULTI-CHANNEL SPEECH ENHANCEMENT}
\name{Wenzhe Liu$^{\star \dagger}$, Andong~Li$^{\star \dagger}$, Chengshi~Zheng$^{\star \dagger}$, Xiaodong~Li$^{\star \dagger}$}
\address{
	$^{\star}$Key Laboratory of Noise and Vibration Research, Institute of Acoustics, Chinese Academy of \\Sciences, Beijing, China\\
	$^{\dagger}$University of Chinese Academy of Sciences, Beijing, China}

\begin{document}
\ninept
\maketitle
\begin{abstract}
 Most deep learning-based multi-channel speech enhancement methods focus on designing a set of beamforming coefficients to directly filter the low signal-to-noise ratio signals received by microphones, which hinders the performance of these approaches. To handle these problems, this paper designs a causal neural beam filter that fully exploits the spatial-spectral information in the beam domain. Specifically, multiple beams are designed to steer towards all directions using a parameterized super-directive beamformer in the first stage. After that, the neural spatial filter is learned by simultaneously modeling the spatial and spectral discriminability of the speech and the interference, so as to extract the desired speech coarsely in the second stage. Finally, to further suppress the interference components especially at low frequencies, a residual estimation module is adopted to refine the output of the second stage. Experimental results demonstrate that the proposed approach outperforms many state-of-the-art multi-channel methods on the generated multi-channel speech dataset based on the DNS-Challenge dataset.
\end{abstract}
\begin{keywords}
Multi-channel speech enhancement, neural beam filter, deep learning, causal
\end{keywords}
\section{Introduction}
\label{sec:intro}
Multi-channel speech enhancement (MCSE) can efficiently suppress directional interference and improve speech quality with beamforming and/or post-filtering~{\cite{overview}}. It has been widely applied as a preprocessor in video conferencing systems, automatic speech recognition (ASR) systems, and smart TVs. The recent breakthrough in deep neural networks (DNNs) has facilitated the research in MCSE, which yields notable performance improvements over conventional statistical beamforming techniques~{\cite{tdcda, sdfcn, avss, wu2020an, gu2019neural, fu2021desnet, mtmvdr, adlmvdr}}. 

Considering the success of DNNs in the single-channel speech enhancement (SCSE) area, a straightforward strategy is to extend the previous SCSE models to extract spatial features either heuristically or implicitly~{\cite{tdcda, sdfcn, avss, wu2020an, gu2019neural, fu2021desnet}}. This paradigm is prone to cause nonlinear speech distortion such as spectral blackhole in low signal-to-noise (SNR) scenarios since the advantage of the spatial filter with microphone-array beamforming is not fully exploited to null the directional interference and suppress the ambient noise~{\cite{mtmvdr, adlmvdr}}. Another category follows the cascade-style regime. To be specific, in the first stage, an SC-based network was adopted to predict the mask of each acoustic channel in parallel, followed by spatial covariance matrices (SCMs) calculation \emph{w.r.t.} speech and noise. In the second stage, traditional beamformers like minimum variance distortionless response (MVDR) or eigenvalue decomposition (GEV) were adopted for spatial filtering~{\cite{mtmvdr, nngev, nnmvdr,gu2021complex}}. These methods have shown their effectiveness in ASR when the latency is not a critical issue, \emph{i.e.}, if it can be up to hundreds of milliseconds. When the latency should be much lower, \emph{i.e.}, $\le$ 20 ms, for many practical applications, such as speech communication, hearing aids, and transparency, these methods may degrade their performance significantly for these low-latency systems. Moreover, the performance heavily depends on the previous mask estimation, which can degrade a lot in complex acoustic scenarios.

As a solution, an intuitive tactic is to enforce the network to directly predict the beamforming weights, which can be done in either time domain~{\cite{luo2019fasnet, luo2020end}} or frequency domain~{\cite{xiao2016deep, adlmvdr, xu2021generalized, ren2021causal}}. Nonetheless, according to the signal theory, the desired beam pattern is required to form its main beam towards the target direction and meanwhile form the null towards the interference direction, which tends to be difficult from the optimization perspective. Moreover, because the network directly determines the weights of the multi-channel input, such a paradigm usually inevitably learns the internal relation of the array geometry, index sequence and the predicted output. In other words, the network itself is not fully decoupled with the inherent information of the array geometry.

\begin{figure*}[t]
	\vspace{-0.3cm}
	\setlength{\abovecaptionskip}{0.235cm}
	\setlength{\belowcaptionskip}{-0.1cm}
	\centering
	\centerline{\includegraphics[width=1.95\columnwidth]{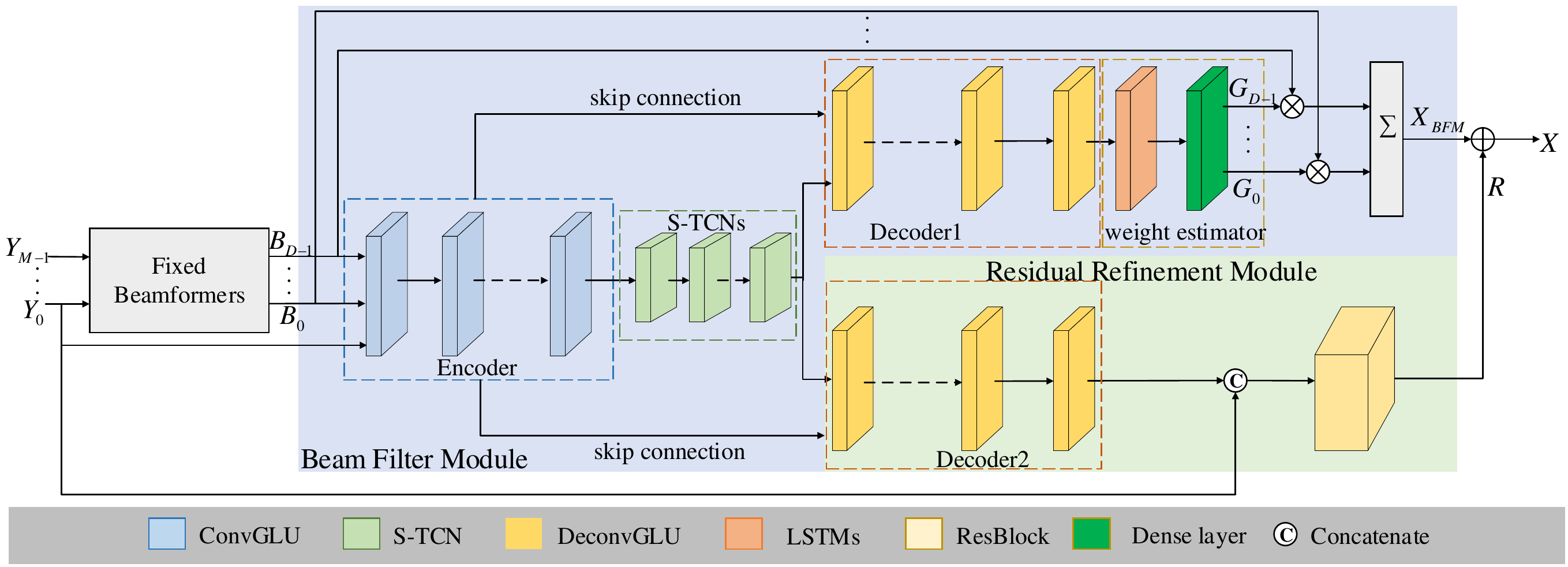}}
	\vspace{-0.4cm}	
	\caption{Overview of the proposed framework. Different modules are highlighted with different colors.}
	\vspace{-0.7cm}	
	\label{fig:overview}
\end{figure*}

In this paper, we design a neural beam filter for real-time multi-channel speech enhancement. In detail, the multi-channel signals are first processed by a set of pre-defined fixed beamformers, \emph{i.e.}, we uniformly sample a beam set with various directions in the space. Then the network is utilized to learn the temporal-spectral-spatial discriminative features of target speech and noise, which aims to generate the bin-level filtering coefficients to automatically weight the beam set. Note that different from previous neural beamformer-based literature where the output weights are applied to multi-channel input signals directly, here the predicted coefficients are to filter the noise component of each pre-generated beam and fuse them. We dub it a neural beam filter to distinguish it from the existing neural beamformer literally. The rationale of such network design logic has three-fold. First, the target signal can be pre-extracted with the fixed beamformer, and the dominant part should exist within at least one or multiple directional beams, which serves as the SNR-improved target priori to guide the subsequent beam fusion process. Second, the interference-dominant beam can be obtained when the beam steers towards the interference direction, which can provide the interference priori for better distinguishment in a spatial-spectral sense. Besides, the target and interference components may co-exist within each beam while their distributions are dynamically changed due to their spectral difference. Therefore, the beam set can be viewed as a reasonable candidate to indicate the spectral and spatial characteristics. As the beam set is only discretely sampled in the space, the information loss tends to arise due to the limited spatial resolution at low frequencies, which causes speech distortion. To this end, a residual branch is designed to refine the fused beam.
We have to emphasize that, although the multi-beam concept is used in both~{\cite{chen2017cracking}} and this study, they are very different as~{\cite{chen2017cracking}} is in essence a parallel single beam enhancement process while the proposed system can be regarded as the filter and fusion process of multi-beam. Experiments conducted on the DNS-Challenge corpus~{\cite{reddy2021icassp}} show that the proposed neural beam filter outperforms previous state-of-the-art (SOTA) baselines.      

The remainder of the paper is organized as follows. We describe the proposed neural beam filter in Section~{\ref{sec:system}}. Experimental setting and results are given in Section~{\ref{sec:expres}} and Section~{\ref{sec:results}}, respectively. Finally, we draw some conclusions in Section~{\ref{sec:conclude}}.

\section{PROPOSED SYSTEM}
\label{sec:system}
\subsection{Forward Stream}
\label{forward-stream}
Fig.~{\ref{fig:overview}} shows the overall diagram of the proposed architecture, which consists of three components, namely fixed beamforming module (FBM), beam filtering module (BFM), and residual refinement module (RRM).

Let us denote the $N$-point short-time Fourier transform (STFT) of $M$-channel input mixtures as $Y_m(t,f)$ $\in\mathbb{C}$, with $m=0,\cdots,M-1$, where $t\in\left\{0,\cdots,T-1\right\}$ and $f\in\left\{0,\cdots,F-1\right\}$ refer to the index of frames and that of frequency bins, respectively.  Considering the symmetry of $Y_m(t,f)$ in frequency, $F = N/2+1$ is chosen throughout this paper. In FBM, the fixed beamformer is employed to sample the space uniformly and obtain multiple beams steering towards different directions, \emph{i.e.}, $B_d\in\mathbb{C}^{T\times F}$, with $d=0,\cdots,D-1$, where $D$ denotes the number of resultant multi-beam. The process is thus given by:
\vspace{-0.2cm}
\begin{gather}
\label{eqn1}
\{B_{d}\}_{d=0,\cdots,D-1} = \mathcal{F}_{FBM}(\{Y_{m}\}_{m=0,\cdots,M-1}; \Phi_{FBM}),
\vspace{-0.2cm}
\end{gather}
where $\mathcal{F}_{FBF}$ is the function of FBM and $\Phi_{FBF}$ denotes the parameter set. From now on, we will omit the subscript $(t,f)$ when no confusion arises. 

We concatenate the beam set along the channel dimension, serving as the input of BFM, \emph{i.e.}, $Cat(B_{0},\cdots,B_{D-1})\in\mathbb{R}^{2D\times T\times F}$. Here $2$ means that both real and imaginary (RI) parts are considered. As muti-beams can represent both spectral and spatial characteristics, BFM is adopted to learn the temporal-spatial-spectral discriminative information between speech and interference and attempt to assign the filter weights $\widehat{G}_{d}\in\mathbb{C}^{T\times F}$ for each beam. It is worth noting that as the beam set is discretely sampled in the space, the information loss tends to arise due to the limited spatial resolution. To alleviate this problem, the complex spectrum of the reference channel is also incorporated into the input and meanwhile, similar to~{\cite{li2021glance}}, the complex residual needs to be estimated with RRM, which aims to compensate for the inherent information loss of the filtered spectrum. This process can be presented as:
\vspace{-0.2cm}
\begin{gather}
\widehat{\mathbf{G}} = \mathcal{F}_{BFM}([\{B_{d}\}_{d=0,\cdots,D-1}, Y_{0}]; \Phi_{BFM}),\\
\widehat{\mathbf{R}} = \mathcal{F}_{RRM} ([\{B_{d}\}_{d=0,\cdots,D-1}, Y_{0}]; \Phi_{RRM}),
\vspace{-0.2cm}
\end{gather}
where $\widehat{\mathbf{G}}\in\mathbb{C}^{D\times T\times F}$ and  $\widehat{\textbf{R}}\in\mathbb{C}^{D\times T\times F}$ denote the complex filter weights and the complex residual estimated by BFM and RRM, respectively. By applying the estimated weights $\{\widehat{G}_{d}\} _{d=0,\cdots,D-1}$ to filter the beams $\{B_{d}\}_{d=0,\cdots,D-1}$ and then summing them along the channel axis, the fused beam $\widehat{X}_{BFM}$ can be obtained by:
\vspace{-0.2cm}
\begin{gather}
	\widehat{X}_{BFM} = \sum_{d}{\widehat{G_{d}}*B_{d}},
	\vspace{-0.2cm}
\end{gather}
where $*$ denotes the complex-valued multiplication operator. We then add the filtered beam and estimated complex residual together to obtain the final output $\widehat{X}$, \emph{i.e},
\vspace{-0.2cm}
\begin{gather}
	\widehat{X} = \widehat{X}_{BFM} + \widehat{R}.
\end{gather}
\vspace{-1.0cm}
\subsection{Fixed Beamforming Module}
In this module, the fixed beamformer is leveraged to transform input multi-channel mixtures into several beams, which steer towards different looking directions and uniformly sample the space. As the fixed beamformer is data-independent, it is robust in adverse environments and has low computational complexity. Moreover, filtering multi-channel mixtures with the fixed beamformer allows our system to be less sensitive to the array geometry and is suitable for arbitrary microphone arrays. In this paper, we choose the super-directivity (SD) beamformer as the default beamformer due to its promising performance in high directivity~{\cite{sd-bf}}. Note that other fixed beamformers can also be adopted, which is out of the scope of the paper. Assume the target directional angle is $\theta_{d}$, the weights of the SD beamformer can be calculated as:
\vspace{-0.2cm}
\begin{gather}
	\mathbf{w}_{d}(f) = \frac{\mathbf{\Gamma}_{nn}^{-1}(f)\mathbf{v}(\theta_{d}, f)}{\mathbf{v}^{H}(\theta_{d},f)\mathbf{\Gamma}_{nn}^{-1}(f)\mathbf{v}(\theta_{d},f)},
\end{gather} 
where $\mathbf{v}(\theta_{d},f)$ is the steering vector, $(\cdot)^{H}$ is the complex transpose operator and $\mathbf{\Gamma}_{nn}(f)$ denotes the covariance matrix of a diffuse noise field with the diagonal loading to control the white noise gain, where we denote it as the parameterized SD beamformer. Note that the diagonal loading level often needs to be chosen carefully to make a good balance between the white noise gain and the array gain~{\cite{pan16reduced}}. In this paper, the diagonal loading level is fixed to 1e-5 and its impact on performance will be studied in the near future. The $(i, j)$-th element of $\mathbf{\Gamma}_{nn}(f)$ represents the coherence between the signals received by two microphones with indices $i$ and $j$ in an isotropic diffuse field, which can be formulated as:
\vspace{-0.2cm}
\begin{equation}
\mathbf{\Gamma}_{nn}^{(i, j)}(f) = \mathrm{sinc}\left(\frac{2\pi f_s f l_{ij} / N}{c}\right),
\end{equation}
where $\mathrm{sinc}(x)=\sin(x)/{x}$ and $l_{ij}$ is the distance between the $i$-th and $j$-th microphones, $c$ is the speed of sound and $f_s$ is the sampling rate.

Define $\mathbf{Y}(t, f) = \{ Y_{0}(t, f), \cdots, Y_{M-1}(t, f) \}$, the output of the $d$-th SD beamformer can be expressed as:
\vspace{-0.2cm}
\begin{gather}
B_{d}(t, f) = \mathbf{w}_{d}^{H}(f)\mathbf{Y}(t, f).
\end{gather} 
\vspace{-0.9cm}
\subsection{Beam Filter Module} 
As shown in Fig.~{\ref{fig:overview}}, the beam filter module (BFM) consists of a causal convolutional encoder-decoder (CED) architecture and a weight estimator. For the encoder, it comprises six two-dimensional (2-D) convolutional gated linear units (ConvGLUs)~{\cite{tan2019learning}}, which consecutively halve the feature size and extract high-level features. Each GLU is followed by instance normalization (IN) and Parameter ReLU (PReLU). The decoder is the mirror version of the encoder except that all the convolution operations are replaced by the deconvolutional version (dubbed DeconvGLU). Similar to~{\cite{li2021two}}, a stack of squeezed temporal convolutional networks (S-TCNs) is inserted as the bottleneck of CED to model the temporal correlations among adjacent frames. After that, in the weight estimator, we simulate the filter generation process, where T-F bin-level filter coefficients are assigned for each beam. To be specific, the output embedding tensor of the decoder is first normalized by layer normalization (LN) and then the LSTM is employed to update the feature frame by frame with ReLU serving as the intermediate nonlinear activation function. The weights $\widehat{\mathbf{G}}$ are obtained after the output linear layer. Then these weights are applied to each beam to obtain the target beam.
\vspace{-0.4cm}
\subsection{Residual Refinement Module}
Because the SD beamformer tends to amplify the white noise to ensure the array gain at low frequencies, the weighted beam output with BFM often contains lots of residual noise components, which need to be further suppressed to improve speech quality. Meanwhile, speech distortion is often introduced due to the mismatch between the main beam steering toward the predefined direction and the true direction of the target speech, which is because the number of the fixed beamformers is limited. To refine the target beam, a residual refinement module (RRM) is proposed, which comprises a decoder module similar to that of BFM and a ResBlock containing three residual blocks, as shown in Fig.~{\ref{fig:overview}}. The output of S-TCNs serves as the input feature of the RRM. After decoding from multiple DeGLUs, the output tensor is concatenated with the original complex spectrum of the reference microphone $Y_{0}$ and is fed to a series of residual blocks. Finally, the complex residual spectrum $\widehat{\mathbf{R}}$ is derived by the output 1$\times$1-conv to reduce the dimensions to 2, and applied to refine the filtered beam output $\widehat{X}_{BFM}$.
%
\vspace{-0.4cm}
\section{EXPEREIMENTAL SETUP}
\label{sec:expres}
\subsection{Datasets}
DNS-Challenge corpus is selected to convolve with multi-channel RIRs to generate multi-channel pairs for experiments. To be specific, the clean clips are randomly sampled from \textit{neutral clean speech} set, which includes about 562 hours speaking by 11,350 speakers. We split it into two parts without overlap, namely for training and testing. For training set, around 20,000 types of noise are selected, with the  duration time about 55 hours~{\cite{li2021two, li2021glance}}. For testing set, four types of unseen noise are chosen, namely babble, factory1, white noises from NOISEX92~{\cite{varga1993assessment}}, and cafe noise from CHiME3~{\cite{barker2015third}}. We generate the RIRs with the image method~{\cite{allen1979image}} based on a uniform linear array with 9 microphones, whose spacing distance is around 4 cm between two neighboring microphones. The room size is sampled from 3$\times$3$\times$2.5 $m^3$ to 10$\times$10$\times$3 $m^3$ and the reverberation time RT60 ranges from 0.05s to 0.7s. The source is randomly located in angle from $0^{\circ}$ to $180^{\circ}$, and the distance between the source and the array center ranges from 0.5 to 3.0 m. The signal-to-interference ratio (SIR) ranges from -6 to 6 dB. Totally, about 80,000 and 4000 multi-channel reverberant noisy mixtures are generated for training and validation. For testing set, five SIRs are set, \emph{i.e.}, $\{-5, -2, 0, 2, 5 \}$ dB and 150 pairs are generated for each case. 

\renewcommand\arraystretch{0.9}
\begin{table*}[t]
	\caption{Objective result comparisons among different causal MCSE models in terms of PESQ and ESTOI for unseen noises. \textbf{BOLD} indicates the best score in each case.}
	\centering
	\scriptsize
	\resizebox{0.94\textwidth}{!}{
		
		\begin{tabular}{c|cccccc|cccccc}
			\hline
			{Metrics} & \multicolumn{6}{c|}{PESQ}  & \multicolumn{6}{c}{ESTOI (in \%)} 
			\\
			\hline
			{SNR (dB)} & -5  &-2 &0 &2 &5&Avg.& -5  &-2 &0 &2 &5&Avg.\\
			\hline
			Unprocessed  &1.36 	&1.44 	&1.58 	&1.70 	&1.90 	&1.60 	&29.99 	&38.90 	&44.72 	&50.66 	&59.22 	&44.70 
			\\
			CTSNet &1.85 	&2.14 	&2.28 	&2.45 	&2.63 	&2.27 	&44.55 	&57.12 	&63.10 	&68.67 	&75.61 	&61.81 
			\\
			GaGNet &1.86 	&2.18 	&2.33 	&2.54 	&2.73 	&2.33 	&46.54 	&59.11 	&64.68 	&70.22 	&77.08 	&63.53 
			\\
			MC-Conv-TasNet  &2.47 	&2.59 	&2.77 	&2.82 	&2.97 	&2.72 	&72.16 	&74.82 	&79.64 	&80.28 	&84.66 	&78.31 
			\\
			MIMO-UNet &2.43 	&2.63 	&2.80 	&2.88 	&2.95 	&2.74 	&66.73 	&72.91 	&78.18 	&79.51 	&82.23 	&75.91 
			\\
			FaSNet-TAC &2.61 	&2.74 	&2.92 	&2.97 	&3.18 	&2.88 	&76.71 	&79.84 &	85.50 	&85.30 	&89.75 	&83.42 
			\\
			\hline
			Proposed-19beams (w/o RRM) &2.95 	&3.11 	&3.30 	&3.36 	&3.57 	&3.26 	&78.53 	&82.34 	&86.89 	&86.87 	&91.10 	&85.15 
			\\
			Proposed-10beams &3.00 	&3.16 	&3.36 	&3.39 	&3.60 	&3.30 &79.27 	&83.00 	&87.35 	&87.25 	&91.52 &85.68\\
			Proposed-19beams &{3.10} 	&{3.24} 	&{3.45} 	&{3.47} 	&3.67 	&{3.39} &{81.54} 	&{84.56} 	&{88.93} 	&88.47	&92.42 &{87.18}
			\\
			Proposed-37beams &\textbf{3.14} 	&\textbf{3.28} 	&\textbf{3.49} 	&\textbf{3.51} 	&\textbf{3.71} 	&\textbf{3.43} &\textbf{82.18} 	&\textbf{85.21} 	&\textbf{89.39} 	&\textbf{88.91} 	&\textbf{92.77} &\textbf{87.69}
			\\
			\hline
	\end{tabular}}
	\vspace{-0.2cm}
	\label{tbl:causal_scores}
\end{table*} 
\vspace{-0.4cm}
\subsection{Baselines}
In this paper, two state-of-the-art (SOTA) single-channel speech enhancement systems namely CTSNet~{\cite{li2021two}} and GaGNet~{\cite{li2021glance}} and three multi-channel speech enhancement systems namely MC-Conv-TasNet~{\cite{luo2019conv}}, FaSNet-TAC~{\cite{luo2019fasnet}}, and MIMO-UNet~{\cite{ren2021causal}} are chosen as the comparative systems. CTSNet is a two-stage framework that ranked 1st in the 2nd DNS Challenge~{\cite{reddy2021icassp}}. As the parallel version of CTSNet, GaGNet is proposed and achieves higher performance. In this paper, we select these two systems to represent the SOTA performance in the monaural speech enhancement area. MC-Conv-TasNet is the multiple-input version of Conv-TasNet which is one of the most effective time-domain speech enhancement and separation models. FaSNet-TAC is an end-to-end filter-and-sum style time-domain multi-channel speech enhancement system, which can achieve better performance than mask-based beamformers. MIMO-UNet is a frequency-domain neural beamformer which is the winner of the INTERSPEECH Far-field Multi-Channel Speech Enhancement Challenge for Video Conferencing~{\cite{rao2021interspeech}}. Note that all the models are set with causal configuration, that is, no future frames are involved into the calculation and inference of current frame.
\subsection{Experiment Setup}
\subsubsection{Network Detail}
In the encoder and decoder part, the kernel size and stride of the 2-D convolution layers are $(2, 3)$ and $(1, 2)$, respectively, and the number of channels remains 64 by default. 3 S-TCNs are adopted, each of which consists of 6 temporal convolutional modules (TCMs) with kernel size and dilation rate being 5 and  $\left(1,2,4,8,16,32\right)$. In the weight estimator module, 2 uni-directional LSTM layers are employed with 64 hidden nodes. In the ResBlock, the kernel size and stride are (2, 3) and (1, 1), respectively.
\vspace{-0.5cm}
\subsubsection{Training Detail}
All the utterances are sampled at 16 kHz. 32 ms Hann window is utilized, with 50\% overlap between adjacent frames. Accordingly, 512-point FFT is utilized, leading to 257-D spectral features. Adam optimizer is applied with the initial learning rate set to 5e-4. If validation loss does not decrease for consecutive two epochs, the learning rate will be halved. All models are trained for 60 epochs.
\vspace{-0.4cm}
\section{RESULTS ANALYSIS}
\label{sec:results}
We choose perceptual evaluation speech quality (PESQ)~{\cite{rix2001perceptual}} and extended short-time objective intelligibility (ESTOI)~{\cite{jensen2016algorithm}} as objective metrics to compare the performance of different models.~{\footnote{Audio samples are available online: https://wenzhe-liu.github.io/NBF/}}
\vspace{-0.4cm}
\subsection{Results Comparison in Objective Metrics}
Table~{\ref{tbl:causal_scores}} shows the objective results of different SE systems. For comparison, the number of beams $D$ is set to 19, \emph{i.e.}, the sampling resolution is 10$^{\circ}$. We evaluate these systems in terms of PESQ and ESTOI. From the table, several observations can be made. First, compared with single-channel speech enhancement approaches such as CTSNet and GaGNet, all the multi-channel speech enhancement systems  yields notable performance improvements consistently thanks to the utilization of spatial information, which indicates that multi-channel information is beneficial for distinguishing speech and interference components. Second, the proposed system outperforms neural beamforming-based approaches by a large margin. For example, compare with FaSNet-TAC, our system achieves 0.51, and 3.76\% improvements in terms of PESQ and ESTOI, respectively. Moreover, our model outperforms MIMO-UNet by 0.65 and 11.27\% in PESQ and ESTOI, respectively. This demonstrates the superiority of filtering the beams over the best neural spatial filters based on frequency and time domains. 
\begin{figure}[t]
	\vspace{-0.2cm}
	\setlength{\abovecaptionskip}{0.235cm}
	\setlength{\belowcaptionskip}{-0.1cm}
	\centering
	\centerline{\includegraphics[width=0.9\columnwidth]{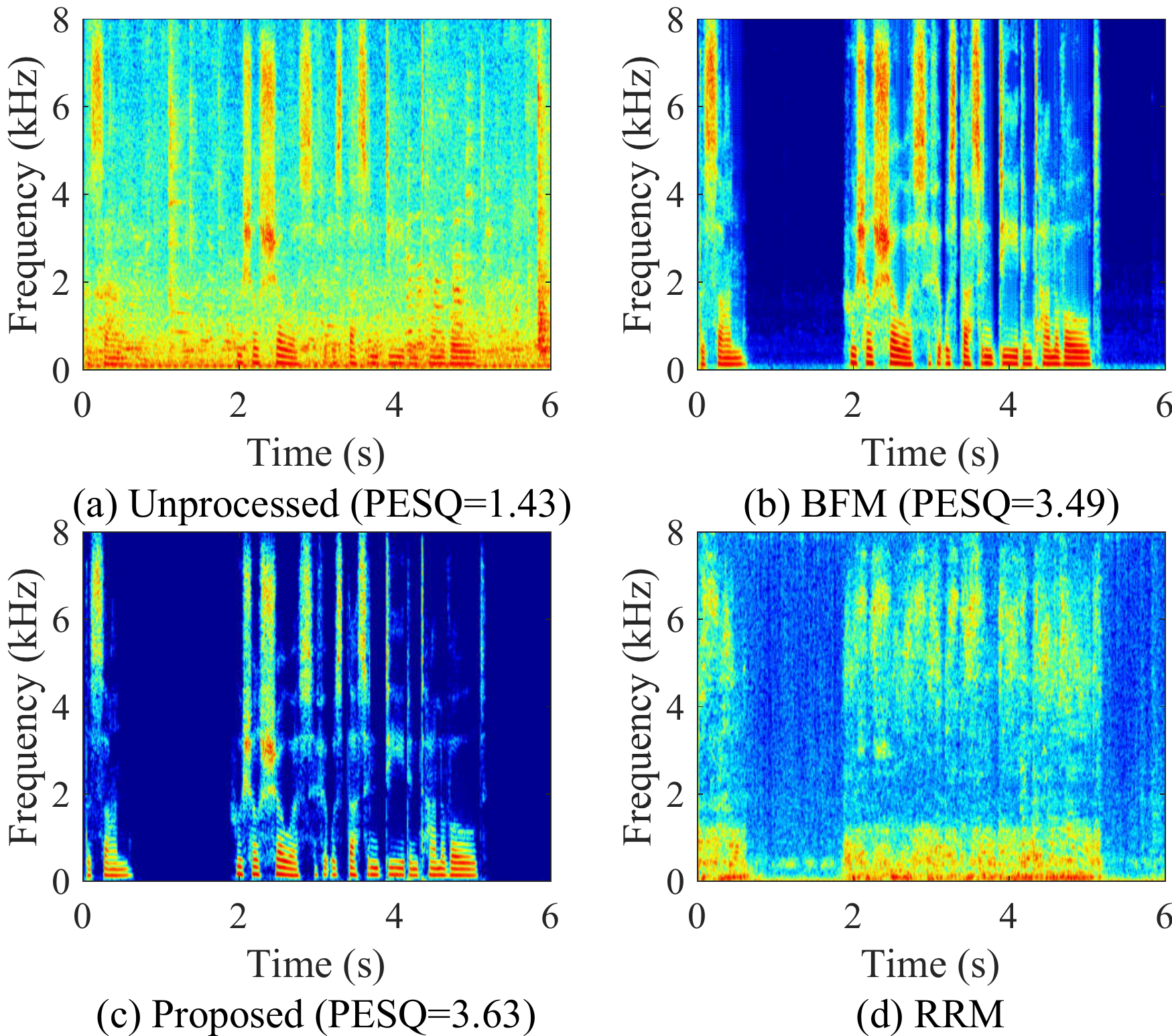}}
	\vspace{-0.4cm}	
	\caption{Visualization results of (a) unprocessed (SIR=0dB), (b) BFM, (c) proposed-19beams, (d) RRM.}
	\label{fig:spec}
	\vspace{-0.45cm}
\end{figure}
\begin{figure}[t]
	\vspace{-0.0cm}
	\centering
	\centerline{\includegraphics[width=0.9\columnwidth]{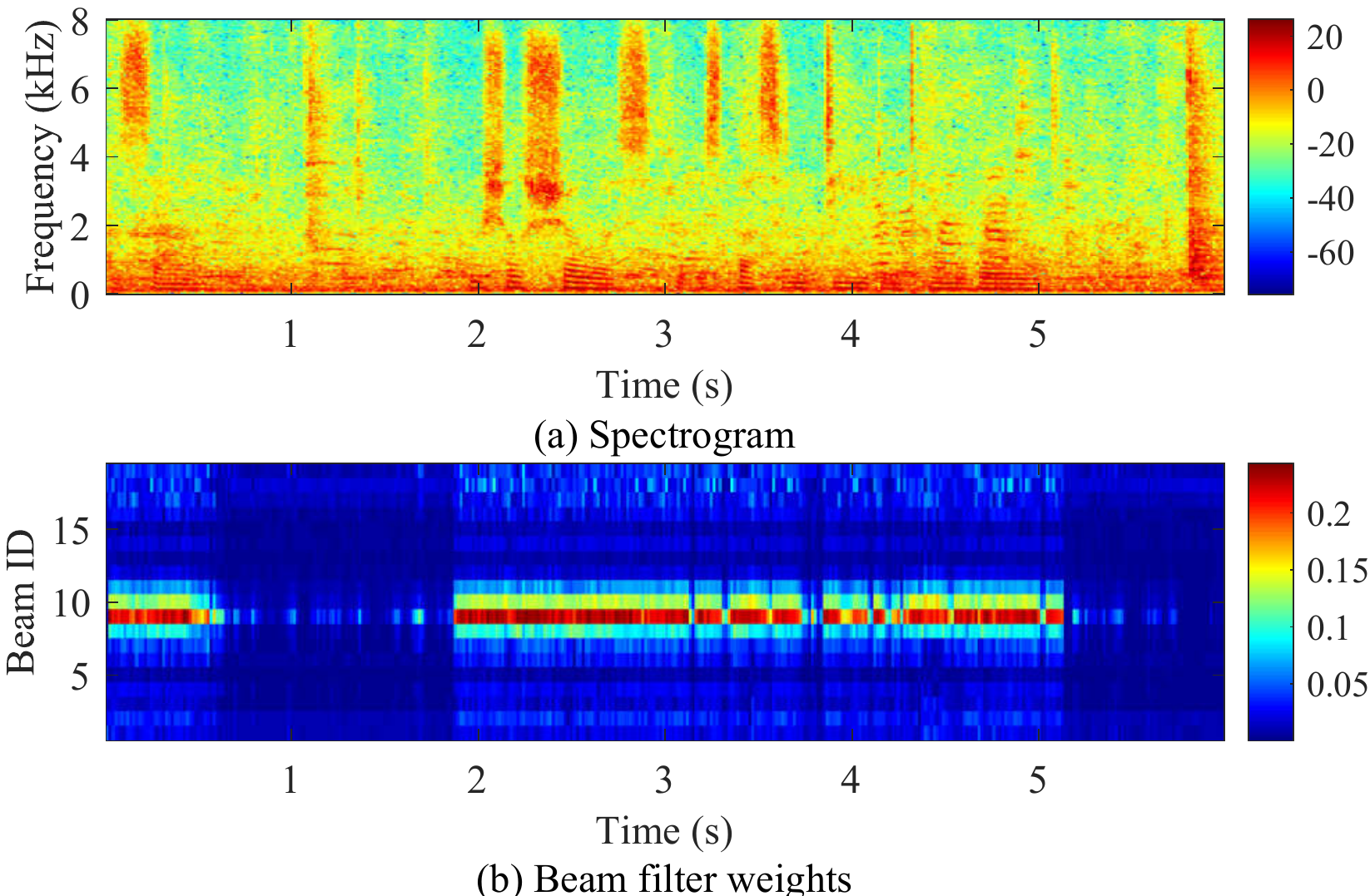}}
	\vspace{-0.3cm}	
	\caption{Visualization results of (a) the spectrogram of input signal, and (b) the estimated filter weights. 
	}
	\label{fig:beam weights}
	\vspace{-0.75cm}
\end{figure}
\vspace{-0.4cm}
\subsection{Ablation Analysis}
We also validate the role of FBM, BFM and RRM. To analyze the effectiveness of FBM, we set another two candidates of $D$, \emph{i.e.}, 10 (20$^\circ$), 37 (5$^\circ$), where 10 ($20^\circ$) means $D = 10$ and each main beam width is about  $20{^\circ}$; 37 ($5{^\circ}$) analogously. It can be seen from Table~{\ref{tbl:causal_scores}} that the performance of the beam neural filter gradually improves with the increase of $D$, which reveals the importance of the FBM. However, the relative performance improvement decreases as the spatial sampling interval becomes progressively smaller although there is still a mismatch between beam pointing and source direction, which indicates that the proposed model is robust to direction mismatch whereas the spatial filter is more sensitive to direction estimation error.
To show the effectiveness of BFM, we visualize the norm of estimated complex weights in Fig.~{\ref{fig:beam weights}}. The input signals are mixed by a speech radiating from $85^{\circ}$ and a Factory1 noise source from $45^{\circ}$. We can find that greater weights are assigned to beams steering toward the surroundings of the target direction, and beams steering to other directions, including those steering toward the interference direction, are given little weights during speaking, while all weights are small in non-speech segments.
Besides, the proposed system with RRM achieves PESQ improvements of 0.13 and ESTOI improvements of 2.03\%. Comparing the visualization results of the model with and without RRM, one can see that the residual noise components are further suppressed at low frequencies and some missing speech components are recovered, which confirms the effectiveness of RRM in the proposed system.
\vspace{-0.4cm}
\section{CONCLUSIONS}
\label{sec:conclude}
In this paper, we propose a causal beam neural filter for real-time multi-channel speech enhancement. It comprises three components, namely FBM, BFM and RRM. Firstly, FBM is adopted to separate the sources from different directions. Then BFM maps filter weights by jointly learning the temporal-spectral-spatial discriminability of speech and interference. Finally, RRM is adopted to refine the weighting beam output. Experimental results show that the proposed system achieves better speech quality and intelligibility over previous spatial neural filters. 

\end{document}